\documentclass[10pt,journal,letterpaper,compsoc]{IEEEtran}

\usepackage{mathptmx}
\usepackage{wrapfig}
\usepackage{graphicx}
\usepackage{times}
\usepackage{color}
\usepackage{soul}
\usepackage{wrapfig}
\usepackage{picins}
\usepackage{caption3}
\usepackage{multirow}
\usepackage{booktabs}

\usepackage[bookmarks,backref=true,linkcolor=black]{hyperref} 
\hypersetup{
  pdfauthor = {},
  pdftitle = {},
  pdfsubject = {},
  pdfkeywords = {},
  colorlinks=true,
  linkcolor= black,
  citecolor= black,
  pageanchor=true,
  urlcolor = black,
  plainpages = false,
  linktocpage
}

\usepackage{multirow}
\usepackage{prettyref}
\newrefformat{fig}{Fig.~\ref{#1}}
\newrefformat{par}{section~\ref{#1}}
\newrefformat{sec}{section~\ref{#1}}
\newrefformat{sub}{section~\ref{#1}}
\newrefformat{table}{table~\ref{#1}}
\newrefformat{alg}{algorithm~\ref{#1}}
\newrefformat{Alg}{Algorithm~\ref{#1}}
\newrefformat{Def}{definition~\ref{#1}}
\newrefformat{Thm}{theorem~\ref{#1}}
\newrefformat{step}{step~\ref{#1}}
\newrefformat{eq}{equation~\ref{#1}}
\newrefformat{pb}{problem~\eqref{#1}}
\def\Eqref Eq:#1:{\eqref{eq:#1}}
\newrefformat{Eq}{Equation~\Eqref#1:}
\newrefformat{appen}{Appendix~\ref{#1}}
\newrefformat{cons}{constraints~\ref{#1}}

\usepackage{amssymb,amsopn,amsmath}
\usepackage{xcolor}
\usepackage{algorithm,algpseudocode}

\newcommand{\CTR}{\textbf{:}}
\newcommand{\E}[1]{\mathbf{#1}}
\newcommand{\IDD}{\operatorname{Id}}

\newcommand{\FPP}[2]{\frac{\partial{#1}}{\partial{#2}}}
\newcommand{\FPPR}[2]{\partial{#1}/\partial{#2}}
\newcommand{\FDD}[2]{\frac{d{#1}}{d{#2}}}

\begin{document}

\title{\Title}
\title{Modelling Developable Ribbons \\Using Ruling Bending Coordinates}
\author{Zherong Pan,
        Jin Huang,~\IEEEmembership{Member,~IEEE,}
        Hujun Bao,~\IEEEmembership{Member,~IEEE}
\IEEEcompsocitemizethanks{
\IEEEcompsocthanksitem The authors are with the State Key Lab of CAD\& CG, Zhejiang University, Hangzhou 310058, China.\protect\\
E-mail: pzr19882@hotmail.com, \{hj, bao\}@cad.zju.edu.cn.}
\thanks{}}

\IEEEcompsoctitleabstractindextext{
\begin{abstract}
This paper presents a new method for modelling the dynamic behaviour of developable ribbons, two dimensional strips with much smaller width than length. Instead of approximating such surface with a general triangle mesh, we characterize it by a set of creases and bending angles across them. This representation allows the developability to be satisfied everywhere while still leaves enough degree of freedom to represent salient global deformation. We show how the potential and kinetic energies can be properly discretized in this configuration space and time integrated in a fully implicit manner. The result is a dynamic simulator with several desirable features: We can model non-trivial deformation using much fewer elements than conventional FEM method. It is stable under extreme deformation, external force or large timestep size. And we can readily handle various user constraints in Euclidean space.
\end{abstract}

}

\maketitle
\IEEEdisplaynotcompsoctitleabstractindextext
\IEEEpeerreviewmaketitle

\section{Introduction}
Developable surfaces are ubiquitous in our daily life. Although their continuous properties have been well understood \cite{do1976differential}, their accurate modelling and discretization is still an open problem. Recently, methods have been proposed in \cite{CGF:CGF1059,CGF:CGF3162} to model these surfaces statically. In this paper, we takes a step further to model the dynamic properties of developable ribbons, a special type of developable surfaces that can be isometrically mapped to two-dimensional strips with much smaller width (latitude dimension) than length (longitude dimension). Developable ribbons have seen a lot of applications for modelling stylish hairs, satin bows or films.

FEM solver is clearly a competitive solution for modelling developable surfaces, using either conforming \cite{Grinspun03discreteshells} or non-conforming triangle meshes \cite{English08animatingdevelopable}. However, none of these methods are geometrically accurate in that their configuration space is not a subset of the true developable shape space. As a result, on conforming meshes large stiffness energies have to be introduced to limit the stretch, which in turn leads to the locking phenomena. On the other hand, the hard length constraints in \cite{english2008animating} on non-conforming meshes greatly limit the timestep size. Moreover, the reconstructed conforming meshes are again not exactly developable and usually suffer from noisy perturbation.

\begin{figure*}[t]
\begin{center}
\includegraphics[width=0.99\textwidth]{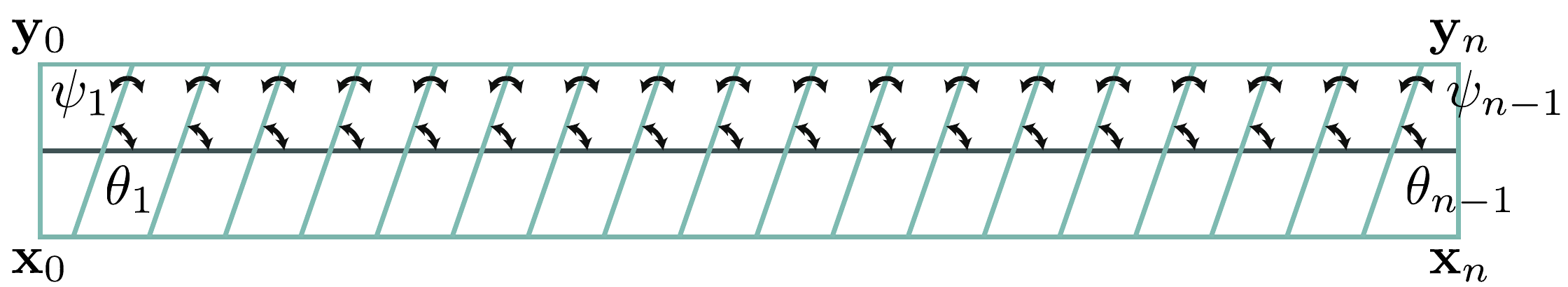}
\end{center}
\caption{Illustration of our configuration space. The centerline of a ribbon is evenly segmented into $n$ elements separated by creases. The angles between these creases and the centerline are $\theta_i$ and the final shape of the ribbon in $\mathcal{R}^3$ is reconstructed by bending along these creases. The corresponding bending angles are $\psi_i$. Given these parameters $\theta_i$ and $\psi_i$, the two ends of each crease, $\E{x}_j$ and $\E{y}_j$, can then be derived analytically.}
\label{fig:Parameter}
\end{figure*}

Key to our dynamic ribbon simulator is a novel configuration space that parameterizes a subset of the developable shape space which is large enough to cover most non-trivial deformations. Specifically, we describe the shape of ribbon by a set of creases along the centerline and bending angles across them, see \prettyref{fig:Parameter}. This method is in direct contrary to previous reduced models such as \cite{Spillmann07corde:cosserat}, where various constrains are introduced to pull the shape towards the true shape space. These constraints usually lead to stability issue or locking phenomena. Instead, our novel representation guarantees that the ribbon can be isometrically mapped to material space. As a result, no additional constraints are needed, making our method stable under large external force or timestep size. Moreover, the whole timestepping scheme can be formulated as a single optimization, which greatly simplifies implementation. We noticed that a similar idea has been exploited in \cite{bertails2006super,casati2013super} for modelling helical rod.

Under this configuration space, we present a proper discretization of the kinetic and potential energies. Our discretization scheme bears several desirable features: First, material space remeshing and world space deformations are modelled uniformly; Energy gradients can be analytically evaluated allowing quasi-newton method to converge efficiently; The $\mathcal{R}^3$ vertex positions $\E{x}_j,\E{y}_j$ are reintroduced as auxiliary variables so that conventional collision handlers can be trivially port to our new formulation. In conclusion, our contributions can be summarized as follows:
\begin{itemize}
\item A novel parameterization of salient global deformations of developable ribbon.
\item A discrete timestepping scheme that can be efficiently integrated in a fully implicit manner.
\item An optimization-based framework for multi-ribbon simulation allowing flexible user constraints and collision resolution.
\end{itemize}

The rest of the paper is organized as follows. After briefly reviewing the related works, we first describe the transfer function between $\mathcal{R}^3$ and our configuration space in \prettyref{sec:Transfer}. We then present our discretization scheme for the kinetic and potential energies in \prettyref{sec:Discrete}. Finally, in \prettyref{sec:Optimization}, we go into some implementation details of our optimization strategy, constrain and collision handling before we conclusion our discussion.

\section{Related Works}
\textbf{Rod Modelling} The theory of elasticity for 1D rod has been established in \cite{cosserat1909theorie}. Later on, various discretization scheme for this model has been developed and applied in robotics \cite{javdani2011modeling}, virtual surgery \cite{CGF:CGF594} and computer animation \cite{spillmann2007c, bertails2006super,bergou2008discrete,bertails2009linear,casati2013super}.

\begin{figure}[h]
\begin{center}
\includegraphics[width=0.49\textwidth]{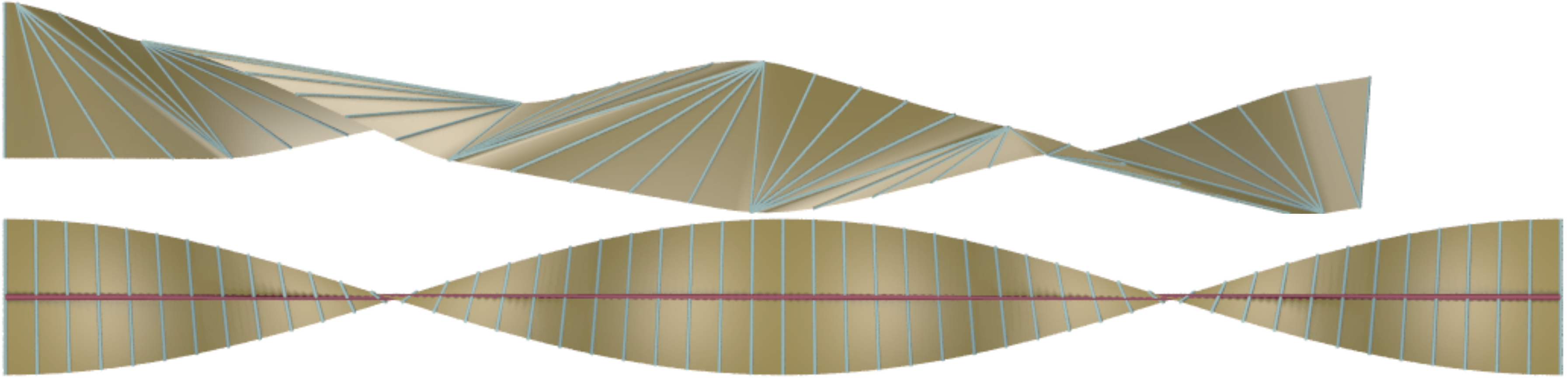}
\end{center}
\caption{\label{fig:compRod} Compared with our method (top), extending a twisted rod (red) along binormal directions (top) doesn't give isometric deformation (area error=$4.7\%$)}
\end{figure}

These discretization schemes fall in two categories: \cite{spillmann2007c} and \cite{bergou2008discrete} adopted a hybrid representation. In their method, the centerline is discretized in $\mathcal{R}^3$ with a frame attached to each segment. This configuration space is not a subset of the true developable shape space so that additional constrains are needed for inextensibility and consistency between the frames and the centerline. Our method is more closely related to \cite{bertails2006super,bergou2008discrete} and \cite{casati2013super} where the configuration space is parameterized solely by the differentials of positions in $\mathcal{R}^3$. These methods share the advantage that no extra constraints are needed. But a reconstruction procedure is required to recover $\mathcal{R}^3$ variables. Despite these similarities, none of them can be directly used to model developable ribbons because their configuration space has only a small intersection with the developable shape space. For example, one may extend a rod along its binormal directions to get a ribbon-like surface but its deformation away from the centerline is not isometric as illustrated in \prettyref{fig:compRod}.

\textbf{Finite Element Shells} Finite element method is another promising alternative for modelling thin shells, see \cite{ciarlet2000theory} for a description of their continuous model. The discrete counterpart has been introduced into the graphics community in \cite{grinspun2003discrete,bridson2003simulation}. But these methods model elastic, instead of developable shells. Later, it is shown in \cite{goldenthal2007efficient} that isometric deformation can be approximated on a conforming mesh by enforcing hard length constraints in a post-projection step. Moreover, this method has the good property that bending energies become quadratic \cite{wardetzky2007discrete}. Although in this work we adopt the dihedral angle based formulates following \cite{grinspun2003discrete,bridson2003simulation}, our bending energies are also quadratic since we treat the bending angles as our generalized coordinates.

However, although a developable surface can be approximated using \cite{goldenthal2007efficient}, a triangular conforming mesh has insufficient degrees of freedom to cover the developable shape space, leading to the so-called locking phenomena. This problem is resolved in \cite{english2008animating} by enforcing the length constraints on a non-conforming mesh. However, \cite{english2008animating} suffers from noisy vertex perturbation when a conforming mesh is reconstruction for rendering and collision resolution. Also, the fast-projection involved in \cite{goldenthal2007efficient,english2008animating} greatly limits the timestep size and stability. Compared with these methods, our formulation allows the same or even higher accuracy with much less elements since no discretization along the latitude direction is needed.

\textbf{Developable Surface Modelling} Our method is also closed related to previous efforts towards static developable surfaces modelling. Among these works, \cite{bo2007geodesic} describes a rectifying developable surface by the envelop of its rectifying planes. However, their method cannot be directly used for dynamic modelling since the centerline is represented by an B\`ezier curve, on which inextensible and other consistency constraints are hard to formulate. Developable surfaces have also been known in the community of architectural design as PQ meshes \cite{liu2006geometric}. Like \cite{spillmann2007c}, developability in their method depends on a set of nonlinear constraints to be satisfied. Finally, our representation is most closely related to \cite{CGF:CGF3162}, where a developable surface is characterized explicitly by creases and bending angles. But we emphasize that, since the correct dynamic behaviour of a developable ribbon heavily depends on the material space crease direction changes, \cite{CGF:CGF3162} cannot be directly used to this end because their crease directions are hard to parameterize.

\section{\label{sec:Transfer}Ruling-Bending Coordinates}
\begin{figure}[h]
\begin{center}
\includegraphics[width=0.49\textwidth]{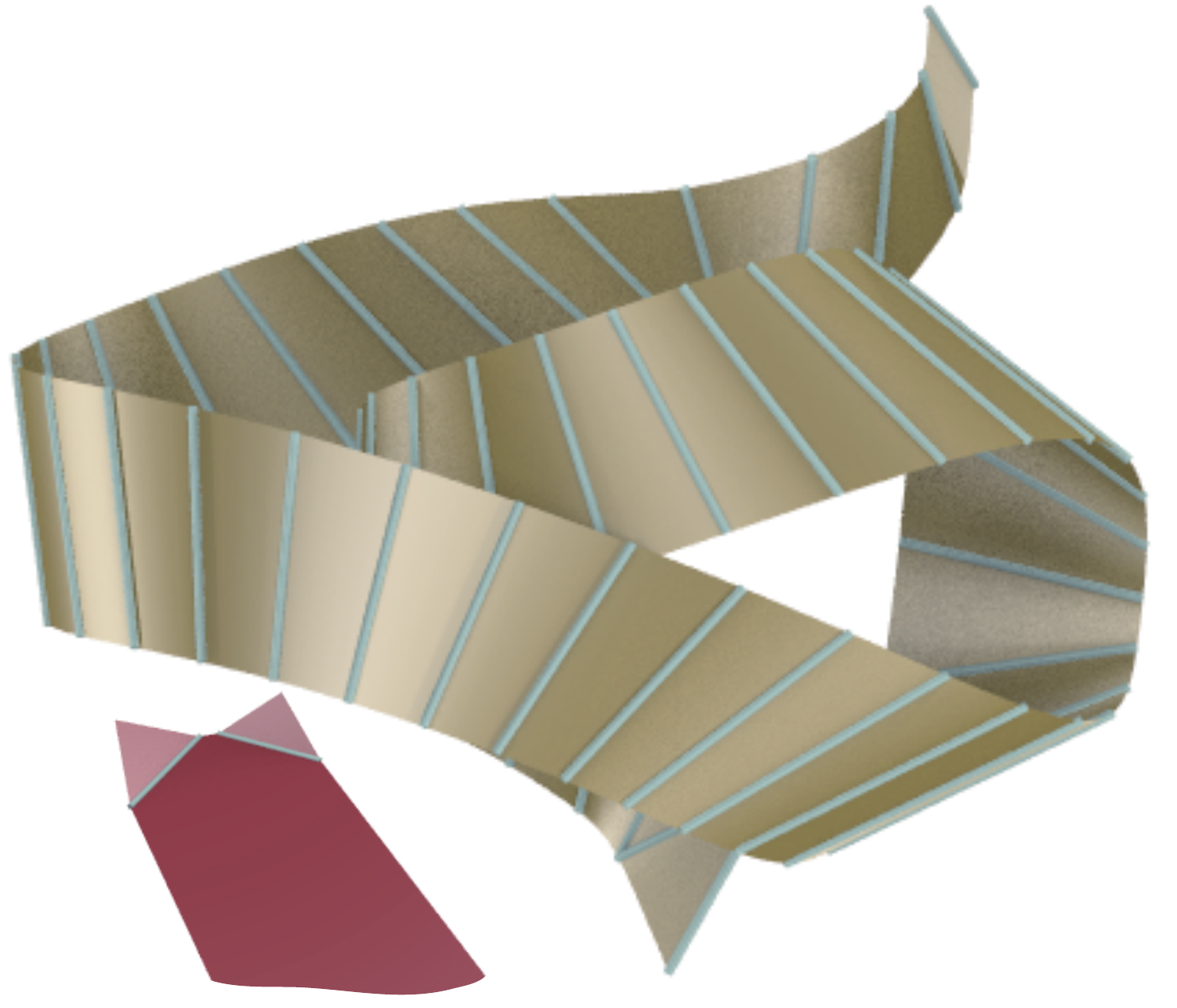}
\end{center}
\caption{Non-trivial global deformations (brown) can be modelled using only creases that cross the centerline. However, some local deformations (red) is excluded from our shape space.}
\label{fig:Deformation}
\end{figure}

One unique property of a developable surface $\E{S}$ is that the Gaussian Curvature is zero everywhere. As a result, each point on $\E{S}$ is attached to a ruling line or crease along which normal is constant and the ribbon is bended by angle $\psi$. As is noted in \cite{CGF:CGF3162}, the world space shape of $\E{S}$ can be characterized by these crease directions and bending angles up to rigid transformation. In our formulation, these two sets of parameters define our configuration space. However, the crease directions for a general developable surface is hard to parameterize. Fortunately, we have observed that for $\E{S}$ with much larger longitude then latitude dimension, a large subset of salient deformations can be modelled with only creases that pass through the ribbon centerline, see \prettyref{fig:Deformation} for an illustration. We can then parameterize our crease direction by $c=\E{tan}(\theta)$, where $\theta$ is the angle between the crease and the centerline.

Specifically, given $\E{S}$ with longitude dimension $l$ and latitude dimension $w$, we first segment its centerline into $n$ elements $\E{E}_{0,\cdots,n-1}$. Then, between any two consecutive elements we introduce creases with directions $c_{1,\cdots,n-1}$ and bending angles $\psi_{1,\cdots,n-1}$. Since we don't allow singularity points, any two consecutive creases cannot intersection, leading to a set of crease constraints:
\begin{eqnarray}
\label{cons:ConsRule}
|c_j-c_{j+1}|\leq\Delta c_{max}\quad
|c_1|\leq\Delta c_{max}\quad
|c_{n-1}|\leq\Delta c_{max},
\end{eqnarray}
where $1\leq j< n-1$ and $\Delta c_{max}=\frac{2l}{wn}$. In practice, we set $\Delta c_{max}=0.95\frac{2l}{wn}$ to avoid degenerate triangles in the reconstructed mesh for collision handling. When these constraints are satisfied, our configuration space is thus parameterized by $<c_i,\psi_i>$.

Like \cite{bertails2006super}, a reconstruction procedure is needed to recover $\mathcal{R}^3$ positions of bottom rim vertices $\E{x}_{0,\cdots,n}$ and top rim vertices $\E{y}_{0,\cdots,n}$. The material space positions of these vertices are:
\begin{eqnarray*}
\bar{\E{x}}_{j}=<\frac{lj}{n}-\frac{wc_j}{2},-\frac{w}{2},0,1>\quad
\bar{\E{y}}_{j}=<\frac{lj}{n}+\frac{wc_j}{2},\frac{w}{2},0,1>,
\end{eqnarray*}
where we need to introduce boundary crease directions $c_0=c_n=0$. Here we used homogeneous coordinates for convenience. Their world space positions $\E{x}_j$ and $\E{y}_j$ can then be derived by applying a series of crease transformations:
\begin{eqnarray}
\label{eq:Recon}
\E{x}_{j}=\left[\Pi_{i=0}^j\E{T}_i\right]\bar{\E{x}}_{j}\quad
\E{y}_{j}=\left[\Pi_{i=0}^j\E{T}_i\right]\bar{\E{y}}_{j},
\end{eqnarray}
where $\E{T}_j$ is a rigid rotation by angle $\psi_j$ along crease $c_j$. In this equation, we again need to introduce boundary value $\E{T}_n=\IDD$. For $\E{T}_0$, if the first segment of the ribbon is fixed, we simply have $\E{T}_0=\IDD$ as well. While if the ribbon is attached to a floating frame, $\E{T}_0$ is a global rigid transformation:
\begin{eqnarray*}
\E{T}_0=\left(\begin{array}{cc}
\E{exp}^\E{w} & \E{t}	\\
\E{0} & 1 \\
\end{array}\right),
\end{eqnarray*}
parameterized by rotation vector $\E{w}$ and translation $\E{t}$. If this is the case, our configuration space is parameterized by the set of variables $<c,\psi,\E{w},\E{t}>$. For numerical optimizaiton, our dynamic simulator heavily depends on an analytical formula for $\nabla\E{x}_j=\FPPR{\E{x}_j}{<c,\psi,\E{w},\E{t}>}$. We leave their derivations to \prettyref{appen:DerivRecon}. $\nabla\E{y}_j$ can be found following the same procedure. Besides, when extra torsional forces are applied on the ribbon, we formulate them as additional constraints on the normal directions in world space. For element $\E{E}_j$, its normal direction $\E{n}_j$ can be calculate by:
\begin{eqnarray}
\label{eq:ReconN}
\E{n}_{j}=\left[\Pi_{i=0}^j\E{T}_i\right]\bar{\E{n}},
\end{eqnarray}
where $\bar{\E{n}}=<0,0,1,0>$. Its derivative $\nabla\E{n}_j$ can be found similarly.

\section{\label{sec:Discrete}Discrete Equation of Motion}
The above configuration space naturally encodes the shape of a globally deformed developable ribbon. To find its motion in temporal domain, we have to discretize the equation of motion. To simplify our presentation, we start from temporal discretization using the Implicit Euler method. It has a simple variational form, which has been exploited in e.g. \cite{martin2011example}:
\begin{eqnarray*}
\E{argmin}_{<c,\psi,\E{w},\E{t}>^{n+1}}\frac{\rho}{2}\left\|\frac{\E{X}^{n+1}-\E{X}^n}{h}-\E{V}^n\right\|_\E{M}^2+V,
\end{eqnarray*}
where $\E{X}$ is position vector assembled from $\E{x}_j,\E{y}_j$ and $\E{V}^n=(\E{X}^n-\E{X}^{n-1})/h$. Here, $V$ denotes the internal or external potential energy terms. Since the ribbon mesh will deform in material space, the mass matrix $\E{M}$ derived from conventional FEM method is dependent on the crease direction $c$. See \prettyref{appen:MassMat} for more details.

In this work, since the configuration space is a subset of the true shape space, no stiffness energies are need to limit stretch and we are left with bending energies to be considered. Since we have the bending angles as an independent variable, bending energies based on dihedral angles \cite{Grinspun03discreteshells,bridson2003simulation} become quadratic in $\psi$ in our case. Specifically, for each crease between $\E{E}_{i-1},\E{E}_{i}$, we introduce:
\begin{eqnarray*}
V_i^{bend}=\int_{D_i}H^2dx\approx\frac{nw(1+c_i^2)\psi_i^2}{l},
\end{eqnarray*} 
where $D_i$ is a diamond element between $\E{E}_{i-1},\E{E}_{i}$ with area $lw/n$. This formula is derived in a similar way to \cite{grinspun2003discrete}. The mean curvature measure of $D_i$ is $\bar{H}_i=w\sqrt{1+c_i^2}\psi_i$ which follows from the tube theory \cite{cohen2003restricted} and the mean curvature is then approximated in a area averaged manner. Although \cite{CGF:CGF3162} adopted a more accurate form, taking the change of mean curvature along ruling into consideration, the difference is insignificant compared with our simplified form.

Other potential terms are discrete version of external forces or soft constraints. For example, the gravitational potential energies are simply: $V^{grav}=-\rho\E{g}^T\E{M}\E{X}^{n+1}$. In addition to these terms, we add a small regularization to resolve the ambiguity of crease directions for a flat ribbon,preferring ruling directions orthogonal to the centerline. The final $V$ is:
\begin{eqnarray*}
V=V^{grav}+\sum_{i=1}^{n-1}\left(\alpha V_i^{bend}+\beta c_i^2\right)+V^{user},
\end{eqnarray*}
where $\alpha$ is the bending stiffness coefficient and $\beta=0.1$ in all our examples. We also added an additional term $V^{user}$ for user controllability.

\begin{figure}[h]
\begin{center}
\includegraphics[width=0.49\textwidth]{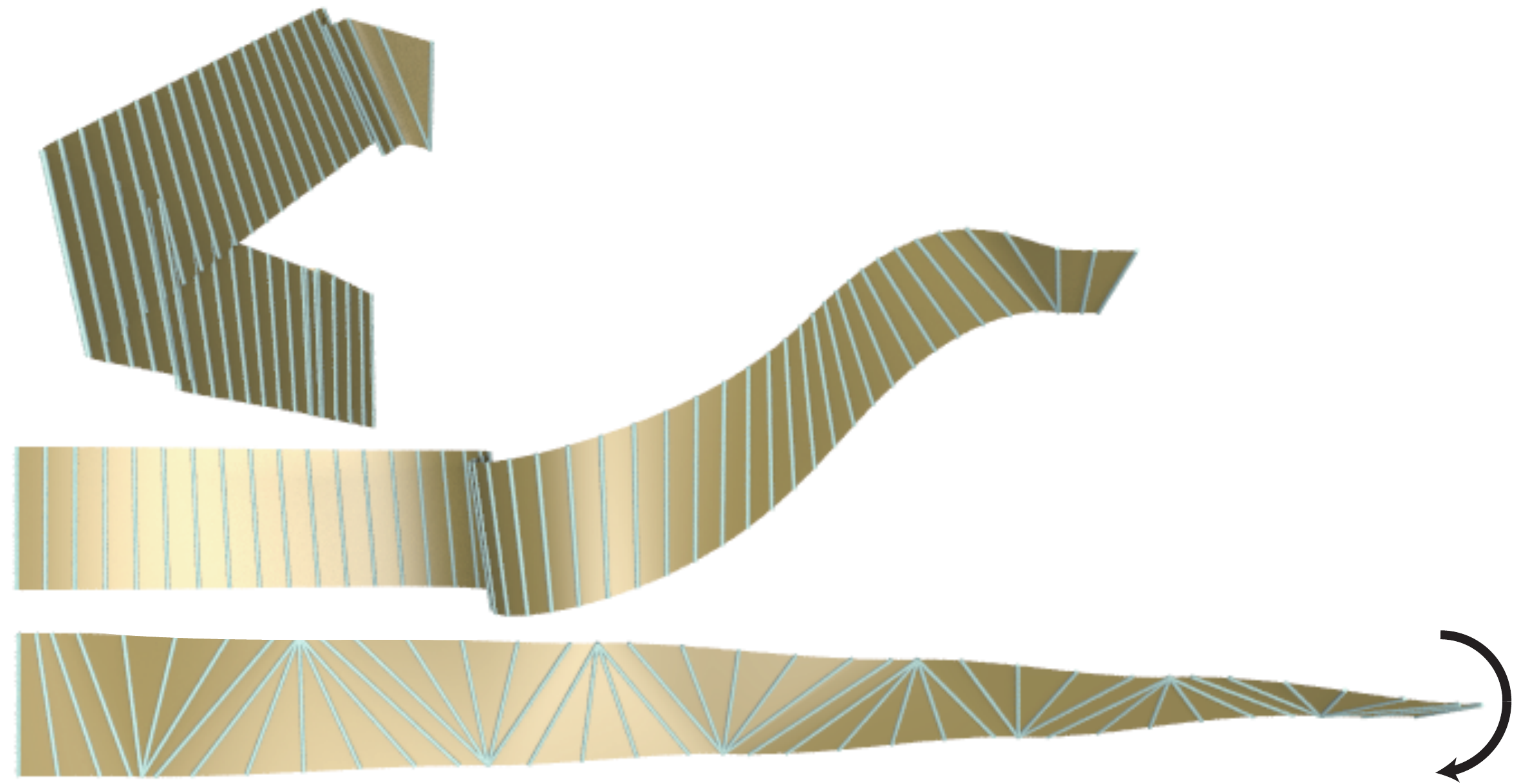}
\end{center}
\caption{A torsional force is applied to a ribbon with one end fixed. Due to our compact configuration space, bending angles alone cannot cover large enough subset of the developable shape space (top). The naive method of updating crease direction once every $5$ frames (middle) still suffer from large deviation from the ground truth (bottom).}
\label{fig:Locking}
\end{figure}

On the other hand, our compact configuration space poses great challenge on spatial discretization of the kinetic term. This is because $<c,\psi>$ encodes material space remeshing (by changing $c$) and world space deformation in a uniform manner. As a result, $\mathcal{R}^3$ positions $\E{X}^{n+1}$ and $\E{X}^{n}$ may not correspond to the same points $\bar{\E{X}}^{n+1}$ and $\bar{\E{X}}^n$ in material space and cannot be subtracted directly. A common practice here is to temporarily fix $\E{c}$ to ensure material space consistency and perform remeshing regularly. Although this strategy has been successfully adopted in conventional FEM shell solver such as \cite{narain2012adaptive}, it fails to work with our method because our configuration space is so compact that bending angles along cannot cover a large enough subset of the true developable shape space, leading to severe locking artifact, see \prettyref{fig:Locking}.

We thus need to introduce a resampling operation $\E{R}$ satisfying:
\begin{eqnarray}
\bar{\E{x}}_j^{n+1}=\E{R}(\bar{\E{x}}_j^{n+1},\bar{\E{X}}^n)\quad\bar{\E{y}}_j^{n+1}=\E{R}(\bar{\E{y}}_j^{n+1},\bar{\E{Y}}^n),
\end{eqnarray}
where $\E{R}$ is assumed to be piecewise linear in its first parameter. In fact, since all vertices lie on the top or bottom rim of the ribbon, $\E{R}$ is a simple 1D-interpolation. We can then define a valid subtraction as: $(\E{X}^{n+1}-\E{R}(\bar{\E{X}}^{n+1},\E{X}^n))/h-\E{R}(\bar{\E{X}}^{n+1},\E{V}^n)$, where $\E{R}(\E{X},\bullet)$ means apply resampling on each $3\times 1$ block of $\E{X}$. Note that in this way we have essentially encoded remeshing and deformation in a single optimization, and our final form of optimization becomes:
\begin{align*}
&\E{argmin}_{<c,\psi,\E{w},\E{t}>^{n+1}}\:f\quad\E{s.t.}\:\prettyref{cons:ConsRule}	\\
&f\triangleq\frac{\rho}{2}\left\|\frac{\E{X}^{n+1}-\E{R}(\bar{\E{X}}^{n+1},\E{X}^n))}{h}-\E{R}(\bar{\E{X}}^{n+1},\E{V}^n)\right\|_\E{M}^2+V.
\end{align*}
We want to emphasize that $\mathcal{R}^3$ positions $\E{X}^n$ here is not required to be reconstructed from our generalized coordinates, allowing the ribbon to temporarily deviate from the developable shape space. This flexibility enables conventional collision handlers to be easily integrated into our framework.

Our solver for this nonlinear optimization is detailed in \prettyref{sec:Optimization}, which heavily relies on an analytical formula for the energy gradient. A naive way of gradient evaluation may follow from \prettyref{appen:DerivRecon} and chain rule. But we show in \prettyref{appen:AdjointMode} that it could be greatly accelerated by evaluating in an adjoint mode.

\section{\label{sec:Optimization} The Solver Framework}
In this section, we present our full-featured multi-ribbon solver framework, including various user constraints or external forces handling and collision resolution. \prettyref{Alg:Outline} provides an outline of our two-step pipeline. In the first substep, the optimization problem is solved for each ribbon in parallel to predicate a desired new configuration where the $\mathcal{R}^3$ vertex positions are reconstructed. The collision handler then finds a corrected collision free $\mathcal{R}^3$ positions in the second substep, temporarily leaving the developable shape space.
\begin{algorithm}[h]
\caption{one timestep of ribbon solver}
\label{Alg:Outline}
\begin{algorithmic}[1]
\Require{a mesh with vertices $<\E{X}^n,\E{V}^n>$}
\Require{a set of user constraints $\E{C}(\E{X})\geq\E{0}$}
\For{each ribbon $r$}\Comment in parallel
\State $<c,\psi,\E{w},\E{t}>_r^{n+1}$=optimize($<\E{X}_r^n$,$\E{V}_r^n>,\E{C}_r$)
\State $<\tilde{\E{X}}_r^{n+1},\tilde{\E{V}}_r^{n+1}>$=reconstruct($<c,\psi,\E{w},\E{t}>_r^{n+1}$)
\EndFor
\State $<\E{X}^{n+1},\E{V}^{n+1}>$=resolveCollision($<\tilde{\E{X}}^{n+1},\tilde{\E{V}}^{n+1}>$)
\end{algorithmic}
\end{algorithm}

For the optimization in our first substep, since the analytical Hessian matrix of \prettyref{eq:Recon} is too costly to evaluate, Newton-type solvers becomes largely unavailable. We thus choose the L-BFGS-B algorithm \cite{byrd1995limited} as our underlying solver, which is wrapped into an Augmented-Lagrangian framework \cite{nocedal2006numerical} to handle nonlinear constraints.

\subsection{Constraints}
Now we discuss several types of constraints supported by our framework. The most important is the non-intersecting ruling constraints. Since these constraints are large in number, we transform them into box constraints by a variable substitution as:
\begin{eqnarray*}
\Delta c_i=c_i-c_{i-1}\quad 2\leq i<n-1,
\end{eqnarray*}
so that they can be handled by the L-BFGS-B algorithm. We are thus left with only two general linear constraints:
\begin{eqnarray*}
|c_1+\sum_{i=2}^{n-1}\Delta c_i|\leq\Delta c_{max},
\end{eqnarray*}
to be handled by the Augmented-Lagrangian framework.

\begin{figure*}[t]
\begin{center}
\includegraphics[width=0.99\textwidth]{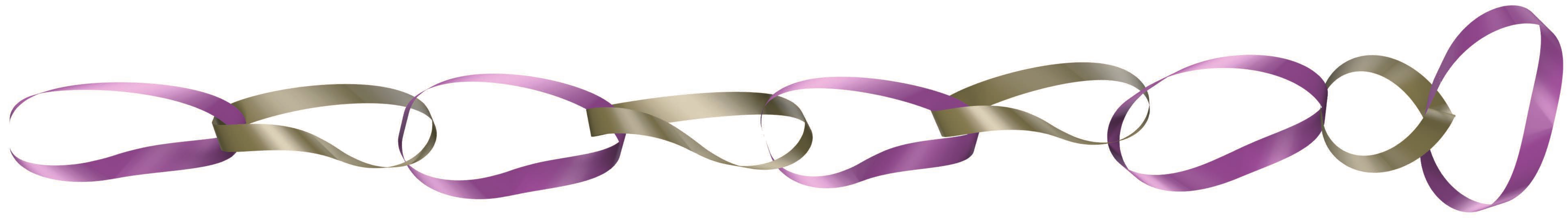}
\end{center}
\caption{Our algorithm can robustly handle nonlinear loop constraints. In this example, an non-orientable loop constraint is assigned to each of the 9 ribbons, forming a Mobius chain.}
\label{fig:Loop}
\end{figure*}

Another common constraint is the loop constraint that requires the two ends of a ribbon to be connected:
\begin{eqnarray*}
\E{x}_0=\E{x}_n\quad\E{y}_0=\E{y}_n\quad\E{n}_0=\E{n}_{n-1}
\end{eqnarray*}
for the orientable case or:
\begin{eqnarray*}
\E{x}_0=\E{y}_n\quad\E{y}_0=\E{x}_n\quad\E{n}_0=-\E{n}_{n-1}
\end{eqnarray*}
for the non-orientable case. These constraints are useful for modelling a ribbon chain or the Mobius band, see \prettyref{fig:Loop}.

Finally, a lot of interesting deformations are resulted from torsional forces which is introduced into our framework as additional normal guiding energies: $V^{user}=K/2\|\E{n}-\E{n}_0\|^2$. Other kinds of common user constraints can be added as in conventional FEM methods. Now we can summarize our optimization substep in \prettyref{alg:Optimize}.
\begin{algorithm}[h]
\caption{optimize($<\E{X}^n$,$\E{V}^n>,\E{C}$)}
\label{alg:Optimize}
\begin{algorithmic}[1]
\State $<c,\psi,\E{w},\E{t}>_0=<c,\psi,\E{w},\E{t}>^n,\lambda=\E{0},\mu=1e^3$
\For{$k=1,2,3,\cdots$}
\State $g=\frac{\mu}{2}\E{min}(\E{C}(\E{X})+\lambda/\mu,\E{0})^2$
\State initial guess $<c_1,\Delta c,\psi,\E{w},\E{t}>_{k-1}=\E{P}<c,\psi,\E{w},\E{t}>_{k-1}$
\State $<c_1,\Delta c,\psi,\E{w},\E{t}>_k=\E{L\_BFGS\_B}(f+g,|\Delta c|\leq\Delta c_{max})$
\State $<c,\psi,\E{w},\E{t}>_k=\E{P}^{-1}<c_1,\Delta c,\psi,\E{w},\E{t}>_k$
\If{$\|<c,\psi,\E{w},\E{t}>_{k-1}-<c,\psi,\E{w},\E{t}>_k\|<1e^{-4}$}
\State return $<c,\psi,\E{w},\E{t}>_k$
\Else
\State $\lambda=\E{min}(\lambda+\mu\E{C}(\E{X}),\E{0})$
\EndIf
\EndFor
\end{algorithmic}
\end{algorithm}

\subsection{Collision Resolution}
One advantage of our formulation is that existing collision detection and resolution methods can be trivially plugged into our framework as a post processor. After a new configuration $<c,\psi,\E{w},\E{t}>^{n+1}$ is returned by our optimizer, a triangle mesh with vertices $\tilde{\E{X}}^{n+1}$ is reconstructed and passed to the collision handler, which in turn finds a closest collision free mesh with vertices: $\E{X}^{n+1}$. Although this mesh may not lie in the developable shape space, its distance to our configuration space is very close in our experiments. Moreover, this developability error will not accumulate because a valid configuration is always recovered by our first substep at next frame.

To work the best with our method, one need to be careful in choosing of underlying continuous collision handler. There are generally two methods for resolving a large amount of continuous collisions: local methods based on randomized impulses \cite{bridson2002robust} and globally coupled methods such as non-rigid impact zone \cite{harmon2008robust}. Since the closeness to a valid configuration is important in our case, we choose to use globally coupled handlers. Specifically, we solve a quadratic programming of the following form to resolve a set of collision constraints $\E{C}_{coll}$ are returned by the detector:
\begin{eqnarray*}
\E{argmin}&&E(\E{X})	\\
\E{s.t.}&&\E{C}_{coll}(\E{X})\geq\E{0},
\end{eqnarray*}
where $E$ is some closeness measure. In the original work \cite{harmon2008robust}, $E$ is simply $E=\frac{1}{2}\|\E{X}^{n+1}-\tilde{\E{X}}^{n+1}\|^2$. This measure leads to diagonal Hessian so that the QP problem can be solved efficiently in its dual form for a very large mesh, especially when the number of constraints are much smaller than number of vertices. Although this measure totally ignores the stiffness between vertices, this choice is appropriate for most cloth animation setups where a small timestep size is used. 

\begin{figure}[h]
\begin{center}
\includegraphics[width=0.49\textwidth]{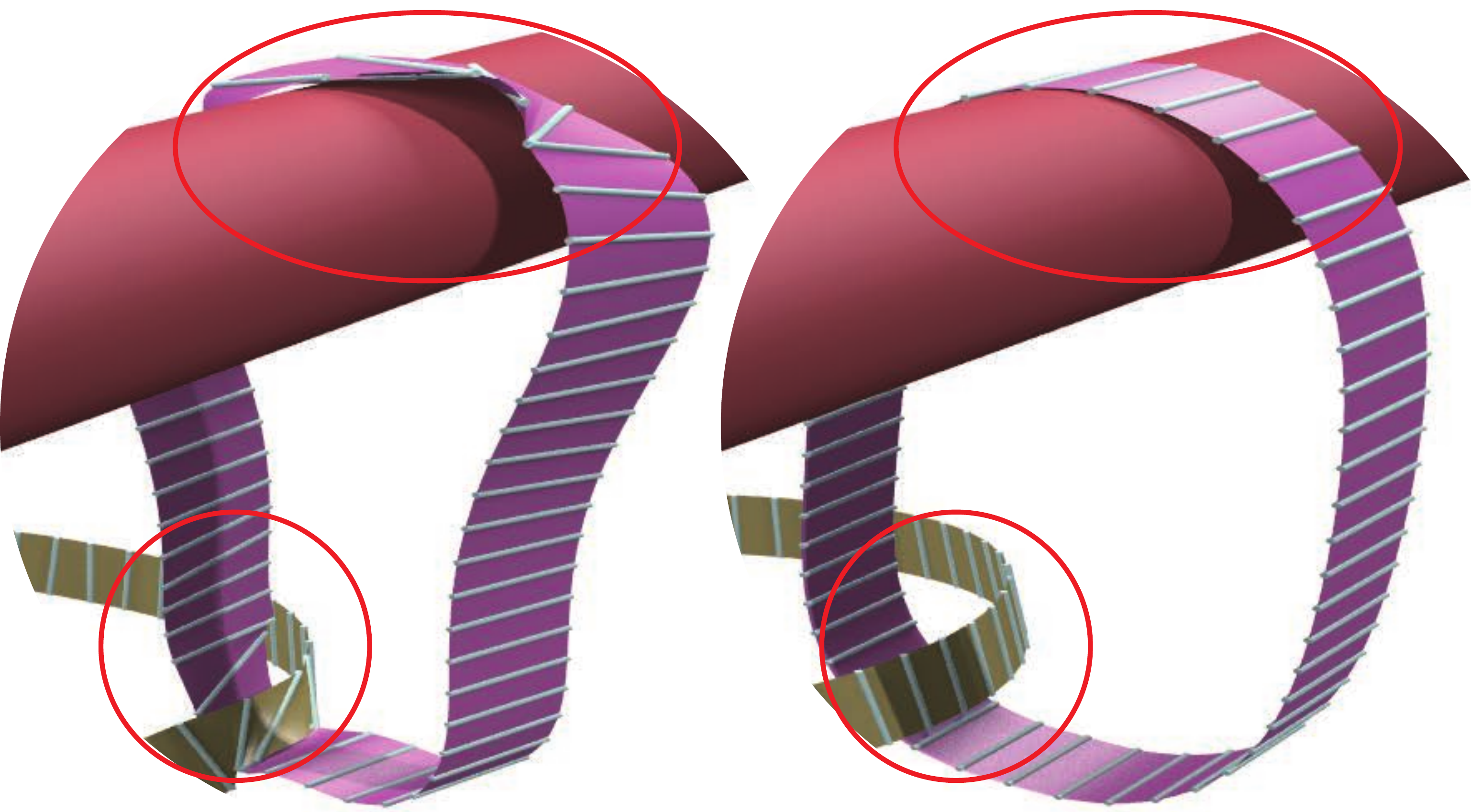}
\end{center}
\caption{A frame from the same animation as bottom \prettyref{fig:ribbonCollision}. Compared with conventional method (left), the collision handler can be greatly stabilized with additional $E_e^{stiff}$ terms (right) under large timestep size.}
\label{fig:CollStiff}
\end{figure}

But in our case, the situation is reversed. Due to our compact representation, the reconstructed mesh is orders of magnitude smaller for comparable results than FEM method. As a result, the number of potential constraints are usually comparable to the number of vertices. On the other hand, since we used fully implicit method, our solver is stable under large timestep. Unfortunately, such large timestep size also makes the collision force stiff. As a result, ignoring internal stiffness in $E(\E{X})$ would lead to large discrepancy between $\tilde{\E{X}}^{n+1}$ and $\E{X}^{n+1}$. This may introduce large developability error and finally cause collision failure due to degenerated or flipped triangles which is illustrated in \prettyref{fig:CollStiff}.

Out of these considerations, we add an quadratic artificial stiffness term to $E(\E{X})$. Specifically, for each edge $\E{e}$ of the reconstructed triangle mesh with vertices $\tilde{\E{x}}_1,\tilde{\E{x}}_2$, we introduce additional energy terms: $E_e^{stiff}(\E{x}_1,\E{x}_2)=\frac{K}{2}\|(\E{x}_1-\E{x}_2)-(\tilde{\E{x}}_1-\tilde{\E{x}}_2)\|^2$, where $K$ is an artificial stiffness coefficient set to $1e^2$ in all our examples. The new QP problem:
\begin{eqnarray*}
\E{argmin}&&E(\E{X})+\sum_e E_e^{stiff}(\E{X})	\\
\E{s.t.}&&\E{C}_{coll}(\E{X})\geq\E{0}
\end{eqnarray*}
is again solved in its dual form using the active set method, where the dual Hessian $\E{C}\E{H}^{-1}\E{C}^T$ is calculated by pre-factorizing the sparse Hessian $\E{H}$ and solve the sparse right hand side $\E{C}^T$. Since our mesh is rather small, the overhead of this solve is neglectable.

\section{Results and Validations}
\begin{figure}[h]
\begin{center}
\includegraphics[width=0.49\textwidth]{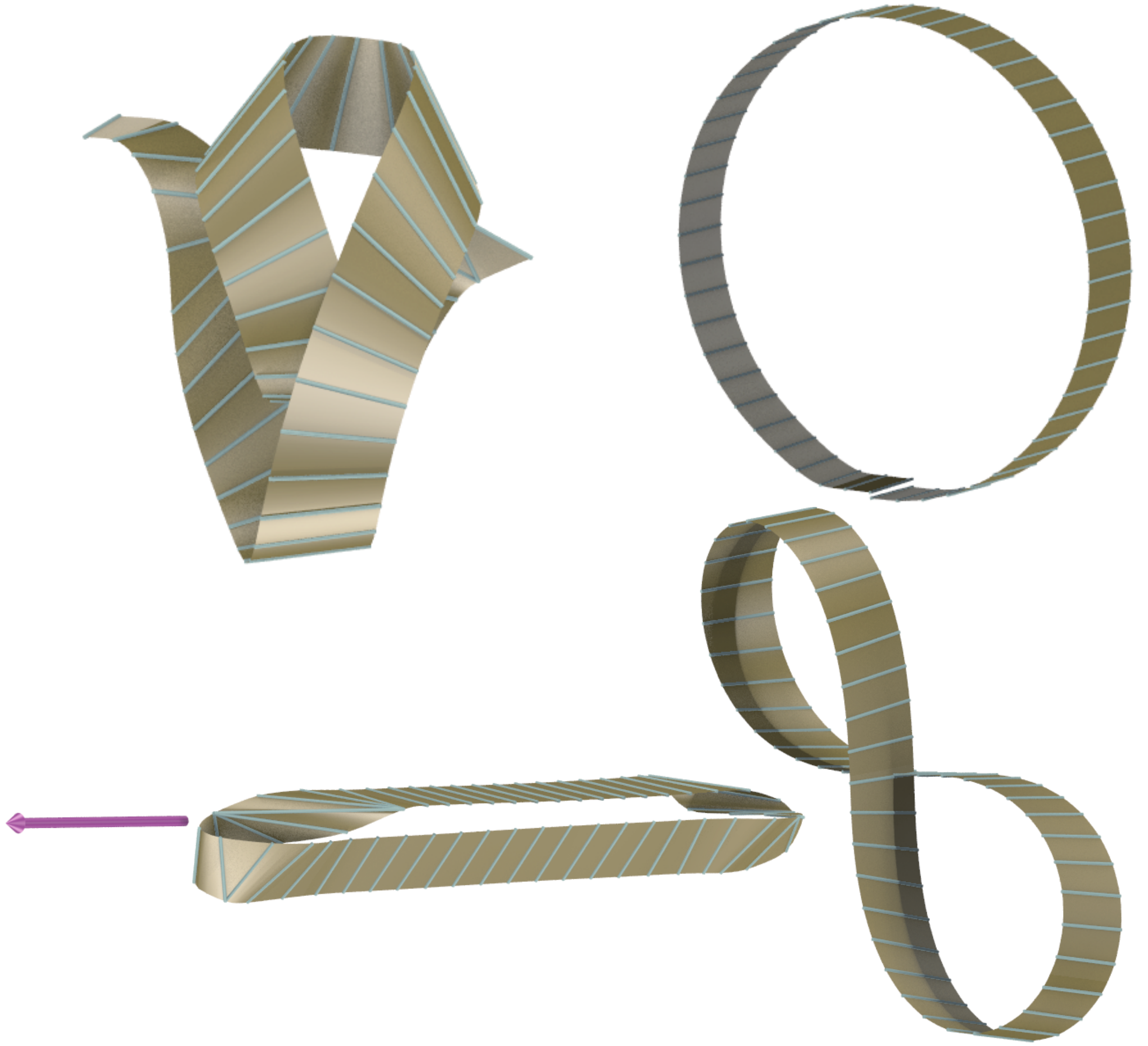}
\end{center}
\caption{Stability test on a ribbon with length=$1m$ and width=$0.05m$ segmented into $50$ elements. Top row: twisting the spherical ribbon by $4\pi$ with strong normal constraints. Bottom row: a looped ribbon under large dragging force.}
\label{fig:LargeForce}
\end{figure}

\begin{figure}[h]
\begin{center}
\includegraphics[width=0.49\textwidth]{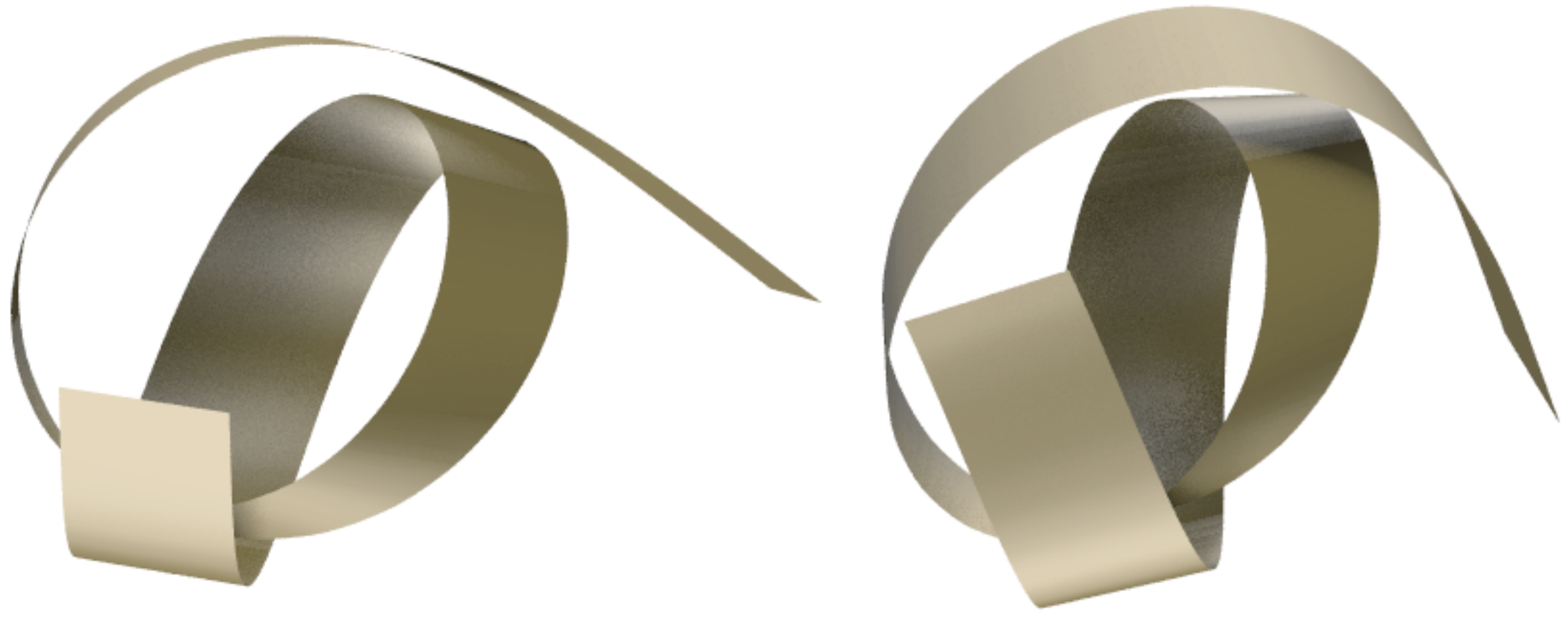}
\end{center}
\caption{Helix unrolling of a ribbon with length=$1m$ and width=$0.05m$ segmented into $50$ elements. Under different timestep size $h=0.01s$ on the left and $h=0.05s$ on the right, our solver finds consistent results.}
\label{fig:Timestep}
\end{figure}

The stability, accuracy and efficiency of our method is evaluated using several benchmark tests, the performance of our solver on all examples is summarized in \prettyref{table:performance}. In a first set of tests illustrated in \prettyref{fig:LargeForce}, we apply large torsional or dragging forces on a ribbon with or without loop constraint. Fortunately, our optimizer presents no performance degradation in both cases.
The stability is also validated in temporal domain. In this subsequent test, we compared two simulated helix unrolling sequences under different timestep size. Our solver finds a reasonably consistent result with $h$ up to $0.05s$, see \prettyref{fig:Timestep}.

\begin{table}[t]
\addtolength{\tabcolsep}{-2pt}
\begin{center}
\begin{tabular}{lcccc}
\toprule
Scene & Ribbon & Opt./[Coll.](sec) & Inner/Outer	\\
\midrule
Extreme Torsion & $50\times 1$ & $0.015$ & $270/1$	\\
Extreme Dragging & $50\times 1$ & $0.03$ & $432/4$	\\
Torsion (ours) & $50\times 1$ & $0.06$ & $110/1$	\\
Torsion (C-FEM) & $50\times 10$ & $0.36$ & N/A	\\
Torsion (NC-FEM) & $50\times 10$ & $3.0$ & N/A	\\
Helix Unrolling (h=0.01) & $50\times 1$ & $0.045$ & $414/1$	\\
Helix Unrolling (h=0.05) & $50\times 1$ & $0.12$ & $439/1$	\\
Ribbon Falling & $150\times 1$ & $2.12/1.2$ & $981/1$	\\
Double Chain & $(50\times 1)\times 18$ & $2.18/2.5$ & $321/5$	\\
Mobius Chain & $(50\times 1)\times 9$ & $2.24/2.1$ & $3417/5$	\\
\bottomrule
\end{tabular}
\end{center}
\caption{\label{table:performance} Time cost for one step of our solver. From left to right: number of ribbon segments in longitude/latitude dimension, average time cost for optimization/collision detection if applicable and number of inner/outer iterations taken by the optimizer. All tests are done on a single desktop computer with dual Intel Xeon E5-2630 CPU and 128Gb memory.}
\vspace{-15px}
\end{table}

\begin{figure}[h]
\begin{center}
\includegraphics[width=0.49\textwidth]{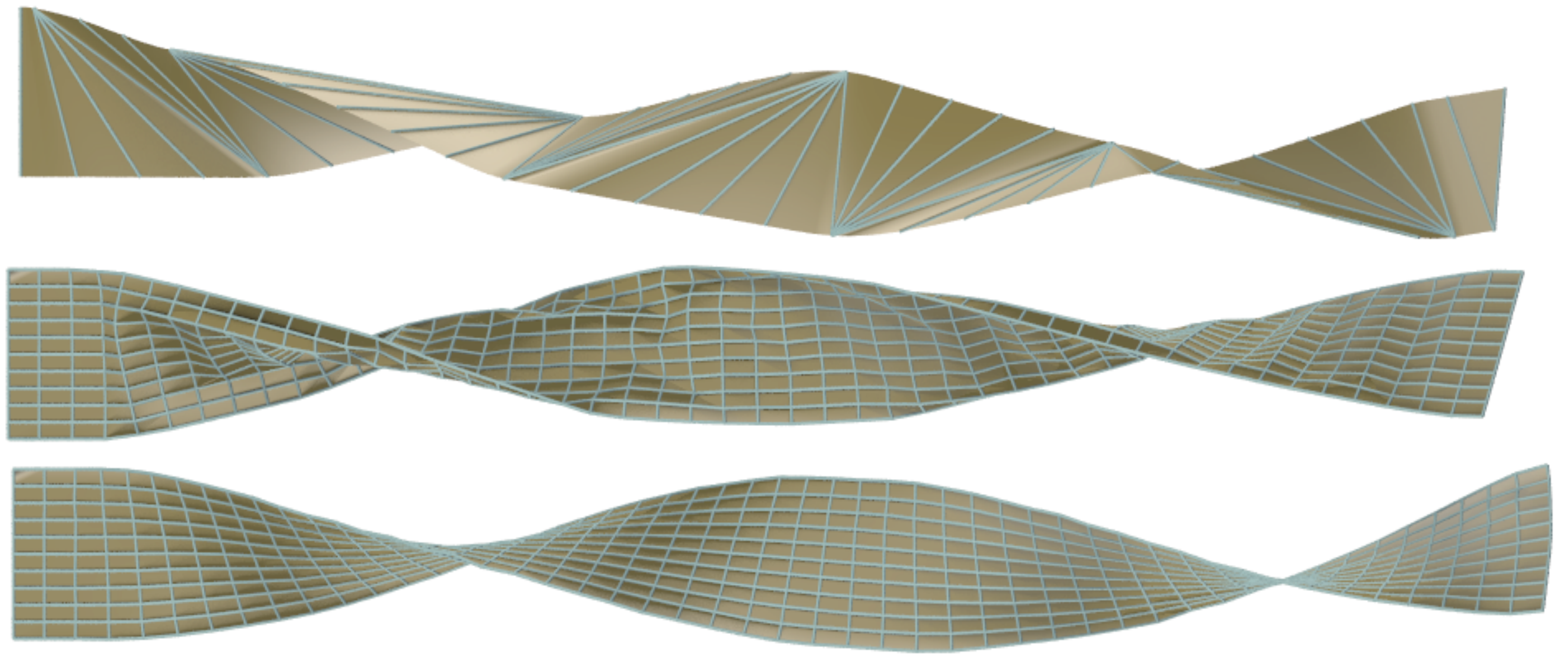}
\end{center}
\caption{Twisting a ribbon with length=$1m$ and width=$0.1m$ segmented into $50$ longitude elements and $10$ latitude elements. From top to bottom: our method, FEM solver on a conforming mesh and FEM solver on a non-conforming mesh.}
\label{fig:Comparison}
\end{figure}

To demonstrate our advantage over previous methods, we compared our solver with conventional FEM methods using (non-)conforming mesh. For the conforming solver, we have to introduce an artificial stiffness term which is set to $1e^8$. According to \prettyref{fig:Comparison} and the accompanying video, the conforming solver would suffer from noisy spatial error due to locking phenomena and the non-conforming solver instead suffers from noisy temporal error due to the conforming reconstruction procedure and additional boundary constraints. By contrast, our method on a much smaller mesh faithfully regenerates the zigzag ruling pattern without any instability in both spatial and temporal domain.

\begin{figure*}[t]
\begin{center}
\includegraphics[width=0.99\textwidth]{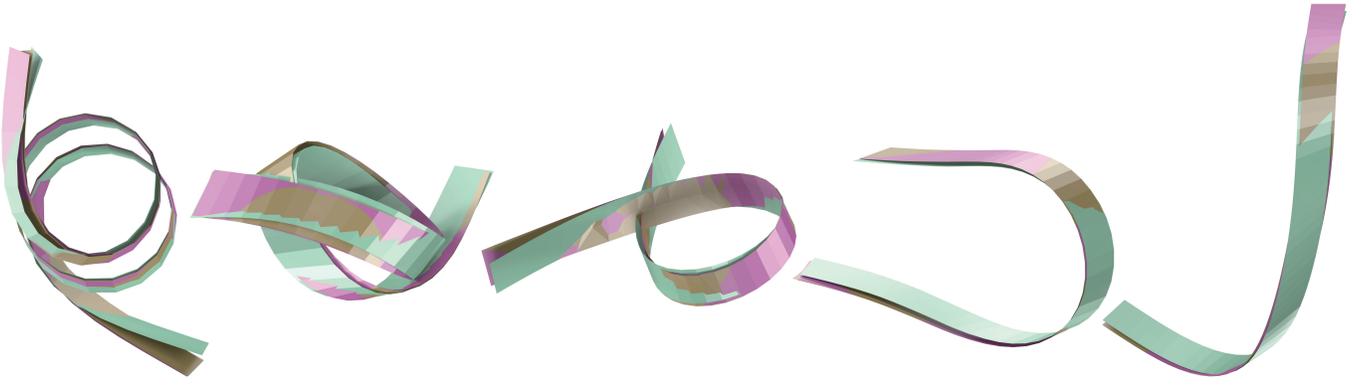}
\end{center}
\caption{The animation of free helix unrolling with $h=0.01s$. Visualization of discrepancy between $\E{Z}_{\textbf{FEM}}^{i+1}$ (red) and $\E{Z}^{i+1}$ (brown) at $t=0s,0.25s,0.5s,0.75s,1s$, from left to right. For reference, $\E{Z}^i$ is also shown in green.}
\label{fig:ComparisonKinetic}
\end{figure*}

\begin{figure}[h]
\begin{center}
\includegraphics[width=0.49\textwidth]{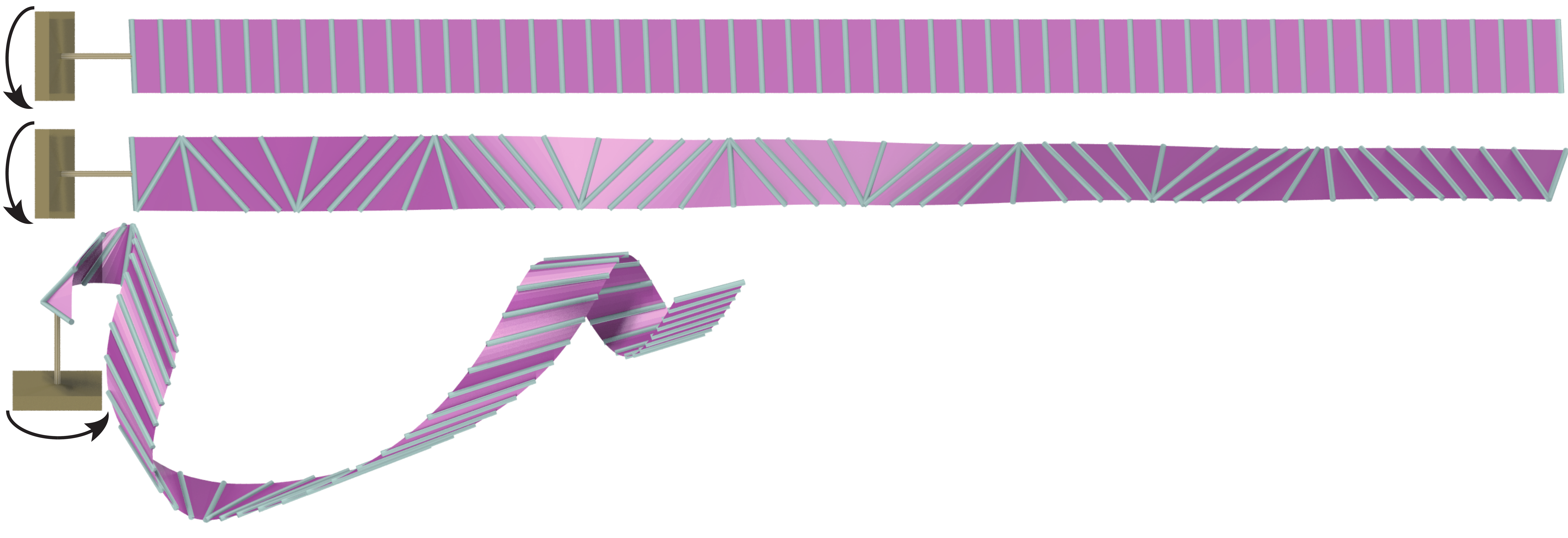}
\end{center}
\caption{A ribbon rotating along the centerline (no gravity). If lumped kinetic energy is used, the ribbon won't even deform since $f=0$ is always achieved at the rest state (top). While our formulation correctly captures the triangular ruling pattern (middle), which finally leads to large deformation due to centrifugal force (bottom).}
\label{fig:Rotate}
\end{figure}

In addition, we investigated the accuracy of our kinetic energy formulation by showing the consistency with conventional FEM methods again. However, as is shown in \prettyref{fig:Comparison}, FEM solver cannot serve as a valid groundtruth. To resolve this contradiction, we observe that our method is equivalent to an conventional FEM solver with dynamic remeshing respecting the ruling direction. In view of this, we work on an animation sequence generated using our method, and for any two consecutive frames $i$ and $i+1$, we insert all vertices to get a combined mesh with material space vertex positions $\bar{\E{Z}}^i\triangleq(\bar{\E{X}}^i,\bar{\E{X}}^{i+1})$. From this mesh with world space vertex positions $\E{Z}^i=(\E{X}^i,\E{R}(\bar{\E{X}}^{i+1},\E{X}^i))$, we perform time integration using FEM method on conforming mesh to predicate a new combined mesh with world space vertex positions $\E{Z}_{\textbf{FEM}}^{i+1}$, which is then compared with $\E{Z}^{i+1}\triangleq(\E{R}(\bar{\E{X}}^i,\E{X}^{i+1}),\E{X}^{i+1})$. This essentially compares the per-frame discrepancy between our method and FEM method with remeshing factored out. The result is visualized in \prettyref{fig:ComparisonKinetic}, which validates the accuracy of our formulation.
Unfortunately, such accuracy is achieved at the cost of an additional resampling operator $\E{R}$. There is however a naive simplification to our kinetic term by lumping all the mass to the centerline, giving a simplified objective energy:
\begin{equation*}
f\triangleq\sum_{i=0}^n\frac{M_i}{2}\left\|(\E{x}_i^{n+1}+\E{y}_i^{n+1}-\E{x}_i^n-\E{y}_i^n)/{2h}-(\dot{\E{x}}_i^n+\dot{\E{y}}_i^n)/2\right\|^2+V,
\end{equation*}
which just takes the average of velocities and positions along the ruling direction for the centerline. Here $M_i$ is the constant lumped mass for the $i$th centerline vertex. Such approximation works well in some cases such as \prettyref{fig:ComparisonKinetic} but may lead to severe artifact elsewhere. An extreme example is shown in \prettyref{fig:Rotate}.

\begin{figure}[h]
\begin{center}
\includegraphics[width=0.49\textwidth]{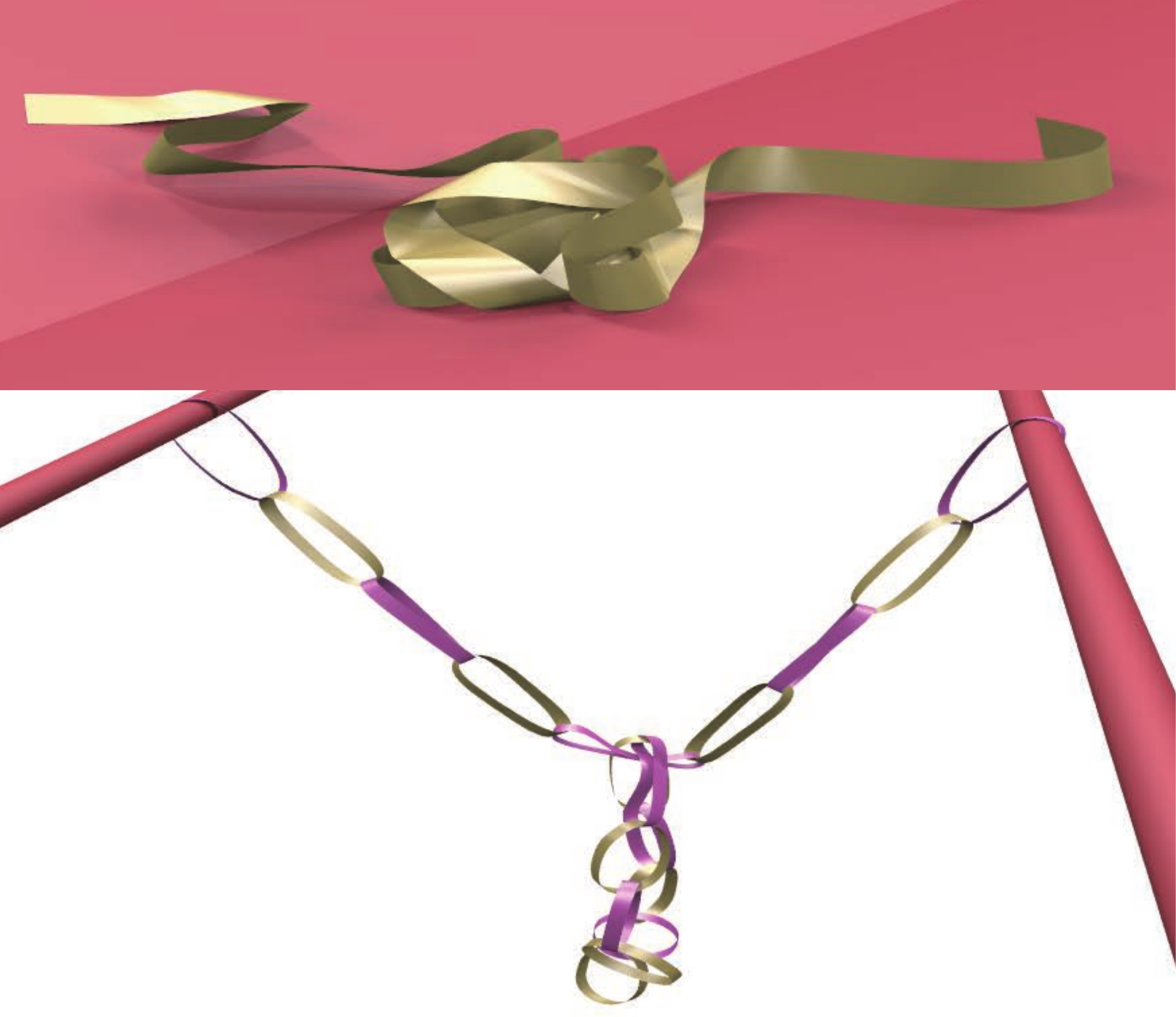}
\end{center}
\caption{Our method handles collision and contact robustly. Top: A long ribbon falling on the ground. Bottom: impact of two ribbon chains with 9 ribbons each.}
\label{fig:ribbonCollision}
\end{figure}

\begin{figure}[h]
\begin{center}
\includegraphics[width=0.49\textwidth]{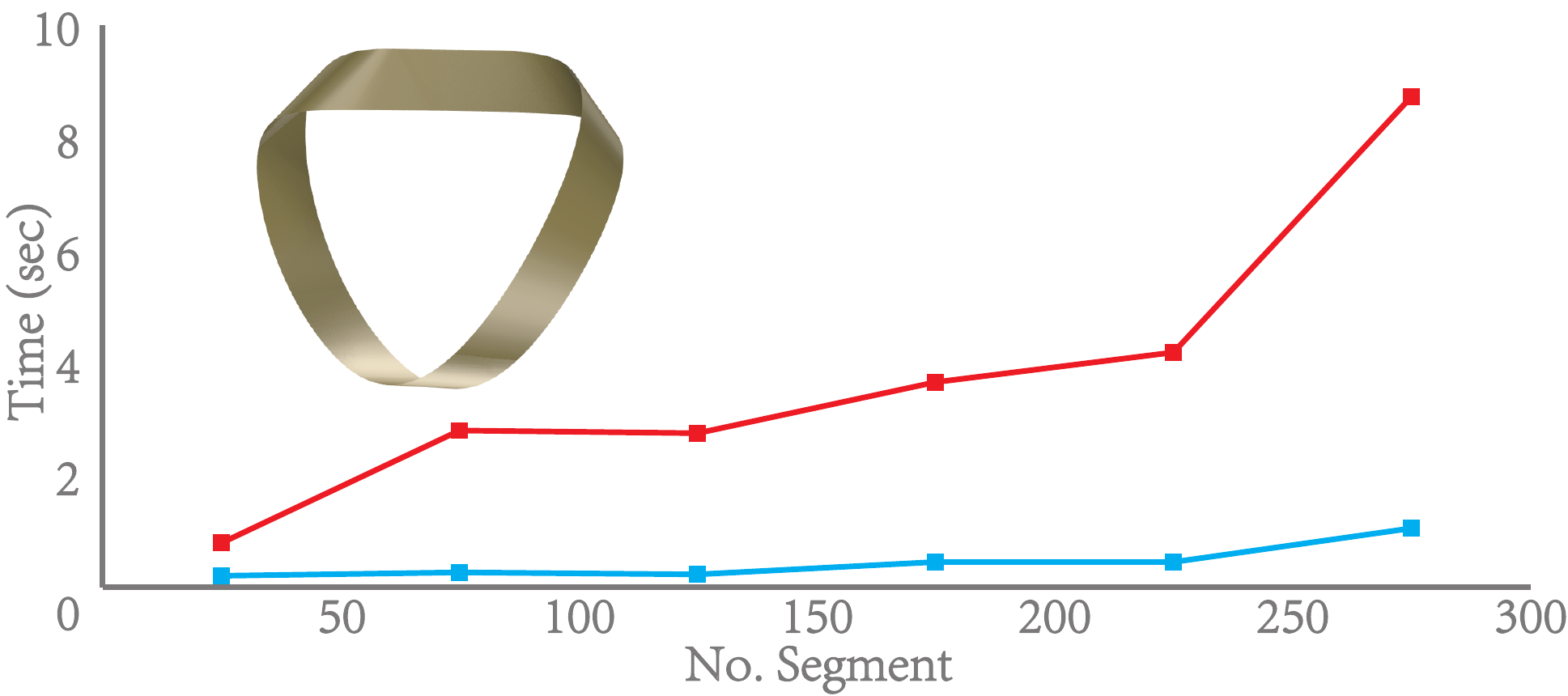}
\end{center}
\caption{Time cost of finding the rest shape of a mobius band for ribbon of different length, using simple chain rule (red) or the adjoint method (blue) energy gradient evaluation.}
\label{fig:perfSeg}
\end{figure}

One major disadvantage of our algorithm is that the performance degenerates with the number of segments according to \prettyref{fig:perfSeg}. The bottleneck of our method is the evaluation of $\nabla\E{x}_j$ and $\nabla\E{y}_j$ which is quadratic in $n$, which can be largely removed using the adjoint method. But longer ribbon won't affect the stability our method. In the falling ribbon example of \prettyref{fig:ribbonCollision}, we used a long ribbon with 150 segments. In this case the overhead of optimization dominates our solver pipeline.

The stability of collision and contact handling is illustrated in \prettyref{fig:ribbonCollision} as well. For both of these examples we used $h=0.01$. Under such large timestep size, our new closeness metric $E$ is an indispensable component especially for the double ribbon chain example. Due to the stiffness of collision forces, conventional non-rigid impact zone solver generates extremely distorted triangles and quickly fails after the first few segments of the chain get in contact.

\section{Conclusion and Discussion}
In conclusion, the paper presents an configuration space that covers a large subset of the shape space of developable ribbon. Based on this configuration space, we develop a optimization based ribbon simulator with flexible user controllability and robust collision handling. Thanks to the compactness of the configuration space, the solver is locking free compared with previous FEM based methods. This enables the solver high fidelity on a very small mesh. Moreover, it presents better stability in both temporal and spatial domain over conventional methods \cite{grinspun2003discrete,english2008animating}.

We also noticed several drawbacks of the method. A clear problem is that our configuration parameters are densely related to vertex positions, which limits the scalability in terms of both time and memory for extremely long chain. The optimizer would also require more gradient evaluations for longer ribbon. To alleviate this problem, it is worth exploring acceleration techniques such as Newton-type optimizer with an approximate Hessian or massive parallelism. Another major drawback is that, as the width of ribbon increase, our configuration space represents a smaller subspace of the true shape space. And sometimes user may want to recover some local deformation. In these cases, locally deformed patches can be reintroduced by coupling the solver to conventional FEM methods or model reduction techniques such as \cite{harmon2013subspace}.

\bibliographystyle{IEEEtran}
\bibliography{refbib}

\begin{appendices}
\section{\label{appen:DerivRecon}Explicit Formula for $\E{x}_j$ and $\nabla\E{x}_j$}
Since $\E{T}_k$ for $1\leq k<n$ is just rigid rotation, they have simple analytical form:
\begin{eqnarray*}
\setlength\arraycolsep{0.05cm}
\E{T}_k=\left(
\begin{array}{cccc}
\frac{\E{cos}(\psi_k)+c_k^2}{c_k^2+1} & \frac{c_k-c_k\E{cos}(\psi_k)}{c_k^2+1} & \frac{-sin(\psi_k)}{\sqrt{c_k^2+1}} & \frac{kl-kl\E{cos}(\psi_k)}{n(c_k^2+1)}	\\

\frac{c_k-c_k\E{cos}(\psi_k)}{c_k^2+1} & \frac{1+c_k^2\E{cos}(\psi_k)}{c_k^2+1} & \frac{c_k\E{sin}(\psi_k)}{\sqrt{c_k^2+1}} & \frac{c_k kl\E{cos}(\psi_k)-c_k kl}{n\sqrt{c_k^2+1}}	\\

\frac{\E{sin}(\psi_k)}{\sqrt{c_k^2+1}} & \frac{-c_k\E{sin}(\psi_k)}{\sqrt{c_k^2+1}} & cos(\psi_k) & \frac{-kl\E{sin}(\psi_k)}{n\sqrt{c_k^2+1}}	\\

0 & 0 & 0 & 1
\end{array}\right),
\end{eqnarray*}
whose partial derivatives $\FPPR{\E{T}_k}{c_k}$ and $\FPPR{\E{T}_k}{w_k}$ can be found using a symbolic software. We omit these here to save space. And for the global rigid transformation, we have:
\begin{eqnarray*}
\FPP{\E{T}_0}{\E{w}_k}=\left(\begin{array}{cc}
\FPP{\E{exp}^\E{w}}{\E{w}_k} & \E{0}	\\
\E{0} & 1 \\
\end{array}\right)\quad
\FPP{\E{T}_0}{\E{t}_k}=\left(\begin{array}{cc}
\E{0} & \E{e}_k	\\
\E{0} & 1 \\
\end{array}\right),
\end{eqnarray*}
where $\FPPR{\E{exp}^\E{w}}{\E{w}_k}$ can be found using the Rodriguez's formula. Now $\nabla\E{x}_j$ can be founding using simply chain rule:
\begin{eqnarray*}
\FPP{\E{x}_{j}}{c_k}&=&\left[\Pi_{i=0}^{k-1}\E{T}_i\right]\FPP{\E{T}_k}{c_k}\left[\Pi_{i=k+1}^j\E{T}_i\right]\bar{\E{x}}_{j}+\left[\Pi_{i=0}^j\E{T}_i\right]\FPP{\bar{\E{x}}_j}{c_k}.	\\
\FPP{\E{x}_{j}}{\psi_k}&=&\left[\Pi_{i=0}^{k-1}\E{T}_i\right]\FPP{\E{T}_k}{\psi_k}\left[\Pi_{i=k+1}^j\E{T}_i\right]\bar{\E{x}}_{j}	\\
\FPP{\E{x}_{j}}{\E{w}_k}&=&\FPP{\E{T}_0}{\E{w}_k}\left[\Pi_{i=1}^j\E{T}_i\right]\bar{\E{x}}_j\quad\FPP{\E{x}_{j}}{\E{t}_k}=\FPP{\E{T}_0}{\E{t}_k}\left[\Pi_{i=1}^j\E{T}_i\right]\bar{\E{x}}_j.
\end{eqnarray*}

\section{\label{appen:MassMat}The Mass Matrix}
Without loss of generality, we use linear shape function for each quad element. In this case, element $\E{E}_i$ would contribute a $12\times12$ mass block of the following form:
\begin{eqnarray*}
\setlength\arraycolsep{0.05cm}
\E{M}_i=\left(\begin{array}{cccc}
\frac{-\Delta c nw^2+4lw}{36n} & \frac{lw}{18n} & \frac{-\Delta c nw^2+4lw}{72n} & \frac{lw}{36n}	\\
\frac{lw}{18n} & \frac{\Delta c nw^2+4lw}{36n} & \frac{lw}{36n} & \frac{\Delta c nw^2+4lw}{72n}	\\
\frac{-\Delta c nw^2+4lw}{72n} & \frac{lw}{36n} & \frac{-\Delta c nw^2+4lw}{36n} & \frac{lw}{18n}	\\
\frac{lw}{36n} & \frac{\Delta c nw^2+4lw}{72n} & \frac{lw}{18n} & \frac{\Delta c nw^2+4lw}{36n}
\end{array}\right)\otimes \IDD^{3\times 3},
\end{eqnarray*}
where $\Delta c=c_{i+1}-c_{i}$.

\section{\label{appen:AdjointMode}Adjoint Mode Gradient Evaluation}
In order to evaluate $\FDD{f}{<c,\psi>}$, we start from the chain rule:
\begin{eqnarray*}
\FDD{f}{<c,\psi>}=\FPP{f}{\E{X}}\FPP{\E{X}}{<c,\psi>}+\FPP{f}{\E{N}}\FPP{\E{N}}{<c,\psi>}+\FPP{f}{<c,\psi>},
\end{eqnarray*}
where the first two terms can then be evaluated in an adjoint mode by exploiting the special structure of \prettyref{eq:Recon} and \prettyref{eq:ReconN}, which is composed of two passes as illustrated in \prettyref{Alg:AdjointAlg}. In this way, the algorithmic complexity is reduced from $\mathcal{O}(n^2)$ to $\mathcal{O}(n)$.
\begin{algorithm}[h]
\caption{Adjoint Gradient Evaluation}
\label{Alg:AdjointAlg}
\begin{algorithmic}[1]
\Require{$<c,\psi>$}
\Ensure{$\FDD{f}{<c,\psi>}$}
\vspace{10px}
\State $\E{T}=\IDD$\Comment forward pass
\For{$j=0,\cdots,n$}
\State $\E{T}=\E{T}\E{T}_j$
\State $\E{x}_j=\E{T}\bar{\E{x}}_j$
\State $\E{y}_j=\E{T}\bar{\E{y}}_j$
\EndFor
\State evaluate $f,\FPP{f}{\E{X}},\FPP{f}{\E{N}},\FPP{f}{<c,\psi>}$
\vspace{10px}
\State $\FDD{f}{<c,\psi>}=\FPP{f}{<c,\psi>}$\Comment backward pass
\State $\E{A}=\E{0}$\Comment adjoint variable
\For{$j=n,\cdots,0$}
\State $\FDD{f}{c_j}=\FDD{f}{c_j}+(\E{T}\FPP{\bar{\E{x}}_j}{c_j})^T\FPP{f}{\E{x}_j}+(\E{T}\FPP{\bar{\E{y}}_j}{c_j})^T\FPP{f}{\E{y}_j}$
\State $\E{T}=\E{T}\E{T}_j^{-1}$
\State $\E{A}=\E{T}_j^{-T}\E{A}\E{T}_{j+1}^T$
\State $\E{A}=\E{A}+\E{T}^T(\FPP{f}{\E{x}_j}\bar{\E{x}}_j^T+\FPP{f}{\E{y}_j}\bar{\E{y}}_j^T+\FPP{f}{\E{n}_j}\bar{\E{n}}^T)$
\State $\FDD{f}{c_j}=\FDD{f}{c_j}+\E{A}\CTR\FPP{\E{T}_j}{c_j}$
\State $\FDD{f}{\psi_j}=\FDD{f}{\psi_j}+\E{A}\CTR\FPP{\E{T}_j}{\psi_j}$
\EndFor
\end{algorithmic}
\end{algorithm}
\end{appendices}





\end{document}